\newcommand{\Tr}{\mathop{\mathrm{Tr}}\nolimits}
\newcommand{\tr}{\mathop{\mathrm{tr}}\nolimits} % Defines the tr operation
\newcommand{\op}[1]{#1}
\newcommand{\Gal}[1]{\mathbb{F}_{#1}}
\newcommand{\prim}{\sigma}

\documentclass[pra,twocolumn,eqsecnum,showpacs,superscriptaddress]{revtex4}
\usepackage{amsfonts,amssymb,bm,graphicx,times,color}

\begin{document}

\title{Mutually unbiased bases and generalized Bell states}

\author{Andrei~B.~Klimov} 
\affiliation{Departamento de F\'{\i}sica,
  Universidad de Guadalajara, 44420~Guadalajara, Jalisco, Mexico}

\author{Denis~Sych}
\affiliation{Max-Planck-Institut f\"ur die Physik des Lichts, G\"{u}nther-Scharowsky-Stra{\ss}e 1, Bau 24, 
91058 Erlangen,  Germany}
\affiliation{Universit\"{a}t Erlangen-N\"{u}rnberg,
Staudtstra{\ss}e 7/B2, 91058 Erlangen, Germany}

\author{Luis~L.~S\'{a}nchez-Soto}
\affiliation{Max-Planck-Institut f\"ur die Physik des Lichts, G\"{u}nther-Scharowsky-Stra{\ss}e 1, Bau 24, 
91058 Erlangen,  Germany}

\author{Gerd~Leuchs}
\affiliation{Max-Planck-Institut f\"ur die Physik des Lichts, G\"{u}nther-Scharowsky-Stra{\ss}e 1, Bau 24, 
91058 Erlangen,  Germany}
\affiliation{Universit\"{a}t Erlangen-N\"{u}rnberg,
Staudtstra{\ss}e 7/B2, 91058 Erlangen, Germany}

\begin{abstract}
  We employ a straightforward relation between mutually unbiased and
  Bell bases to extend the latter in terms of a direct construction
  for the former. We analyze in detail the properties of these new
  generalized Bell states, showing that they constitute an appropriate
  tool for testing entanglement in bipartite multiqudit systems.
\end{abstract}


\maketitle

\section{Introduction}

Entanglement is probably the most intriguing feature of the quantum
world, the hallmark of correlations that delimits the boundary between
classical and quantum behavior. Although some amazing aspects of this
phenomenon were already noticed by Schr\"{o}dinger in the early stages
of quantum theory~\cite{Schrodinger:1935}, it was not until quite
recently that it attracted a considerable attention as a crucial
resource for quantum information processing~\cite{Nielsen:2000}.
  
The simplest instance of entanglement is most clearly illustrated 
by the maximally entangled states between a pair of qubits (known 
as Bell states), whose properties can be found in many
textbooks~\cite{Peres:1993}. Despite their simplicity, they 
are of utmost importance for the analysis of many
experiments~\cite{Wei:2007}.

In consequence, as any sound concept, Bell states deserve an
appropriate generalization. However, this is a touchy business, 
since thoughtful notions for a pair of qubits, may become fuzzy 
for more complex systems.  There are two sensible ways to proceed: 
the first, is to investigate multipartite entanglement of qubits.  
While the standard Bell basis defines (for pure states) a natural 
unit of entanglement, it has recently become clear that for qubits 
shared by more parties there is a rich phenomenology of entangled
states~\cite{Dur:2000,Durr:2000b,Acin:2000,Briegel:2001,Verstraete:2002,
Rigolin:2006,Facchi:2008}.

The second possibility involves examining bipartite entanglement
between two multidimensional
systems~\cite{Bechmann:2000,Bourennane:2001,Cerf:2002,Sych:2004,Sych:2009}.
Again there is no unique way of looking at the problem, and different
definitions focus on different aspects and capture different features
of this quantum phenomenon.

We wish to approach this subject from a new perspective: our starting
point is the notion of mutually unbiased bases (MUBs), which emerged 
in the seminal work of Schwinger~\cite{Schwinger:1960} and it has turned 
into a cornerstone of quantum information, mainly due to the elegant work 
of Wootters and coworkers~\cite{Wootters:1987,Wootters:1989,
Wootters:2004,Gibbons:2004b,Wootters:2006}. Since MUBs contain
complete  single-system information and Bells bases about bipartite
entanglement, one is led to look for a relation between them.

In this paper we confirm such a relation for qudits~\cite{Planat:2005}
and take advantage of the well-established MUB machinery (in prime
power dimensions) to propose a straightforward generalization of Bell
states for any dimension. The resulting bases are analyzed in detail,
paying special attention to their symmetry properties. In view of the
results, we conclude that these states constitute an ideal instrument
to analyze bipartite multiqudit systems.

\section{Bipartite qudit systems}

\subsection{Mutually unbiased bases for qudits}

We start by considering a qudit, which lives in a Hilbert space
$\mathcal{H}_{d}$, whose dimension $d$ is assumed for now to be a
prime number. The different outcomes of a maximal test constitute an
orthogonal basis of $\mathcal{H}_{d}$.  One can also look for other
orthogonal bases that, in addition, are ``as different as possible''.

To formalize this idea, we suppose we have a number of orthonormal
bases described by vectors $ | \psi_{\ell}^{n} \rangle $, where $\ell$
($\ell = 0, 1, \ldots, d-1$) labels the vectors in the $n$th basis.
These are MUBs if each state of one basis gives rise to the same
probabilities when measured with respect to other basis:
\begin{equation}
  \label{eq:MUB2}
  | \langle \psi_{\ell^{\prime}}^{n^{\prime}} | \psi_{\ell}^{n} \rangle |^{2}  = 
  \frac{1}{d} \, , \qquad n \neq n^\prime \, . 
\end{equation}
Equivalently, this can be concisely reformulated as
\begin{equation}
  \label{eq:MUB3}
  | \langle \psi_{\ell^{\prime}}^{n^{\prime}} | \psi_{\ell}^{n} \rangle |^{2}  =
  \delta_{\ell \ell^{\prime}} \delta_{n n^{\prime}} + 
  \frac{1}{d}  ( 1-\delta_{n n^{\prime }} ) \, . 
\end{equation}
Note in passing that the Hermitian product of two MUBs is then 
a  generalized Hadamard matrix, i.e., a unitary matrix whose
entries all have the same absolute value~\cite{Bengtsson:2007}.

If one wants to determine the state of a system, given only a limited
supply of copies, the optimal strategy is to perform
measurements with respect to MUBs. They have also been used
in cryptographic protocols~\cite{Asplund:2001}, due to the complete
uncertainty about the outcome of a measurement in some basis after the
preparation of the system in another, if the bases are mutually
unbiased. MUBs are also important for quantum error correction
codes~\cite{Gottesman:1996,Calderbank:1997} and in quantum game
theory~\cite{Englert:2001,Aravind:2003,Paz:2005,Kimura:2006}.

The maximum number of MUBs can be at most
$d+1$~\cite{Ivanovic:1981}. Actually, it is known that if $d$ is prime
or power of prime (which is precisely our case), the maximal number of
MUBs can be achieved.

Unbiasedness also applies to measurements: two nondegenerate tests 
are mutually unbiased if the bases formed by their eigenstates are MUBs.
For example, the measurements of the components of a qubit along $x$,
$y$, and $z$ axes are all unbiased. It is also obvious that for these
finite quantum systems unbiasedness is tantamount of
complementarity~\cite{Kraus:1987,Lawrence:2002}.

The construction of MUBs is closely related to the possibility of
finding of $d+1$ disjoint classes, each one having $d-1$ commuting
operators, so that the corresponding eigenstates form sets of
MUBs~\cite{Bandyopadhyay:2002}. Different explicit methods in 
prime power dimensions have been suggested in a number of
recent papers~\cite{Klappenecker:2004,Lawrence:2004,Pittenger:2005,
Wocjan:2005,Durt:2005,Klimov:2007}, but we follow here the one 
introduced in Ref.~\cite{Klimov:2005}, since it is especially germane 
for our purposes. 

First, we choose a computational basis $| \ell  \rangle $ in 
$\mathcal{H}_{d}$ and introduce the basic operators
\begin{equation}
  \label{CC}
  X | \ell \rangle =  | \ell + 1 \rangle \, ,
  \qquad \qquad
  Z | \ell \rangle  =  \omega (\ell) | \ell \rangle \, ,
\end{equation}
where addition and multiplication must be understood modulo $d$ and,
for simplicity, we employ the notation 
\begin{equation}
\omega ( \ell ) =
\omega^{\ell} = \exp (i 2 \pi \ell /d) \, ,
\end{equation} 
$\omega = \exp( i 2\pi/d)$ being a $d$th root of the unity.  These
operators $X$ and $Z$, which are generalizations of the Pauli
matrices, were studied long ago by Weil~\cite{Weil:1964}.  They
generate a group under multiplication known as the generalized Pauli
group and obey $Z X = \omega \, X Z$, which is the finite-dimensional
version of the Weyl form of the commutation
relations~\cite{Putnam:1987}.

We consider the following sets of operators:
\begin{equation}
  \label{SCop}
  \tilde{\op{\Lambda}} ( m ) = \op{X}^{m}  \, ,
  \qquad 
  \op{\Lambda} (m, n) =  \op{Z}^{m} \op{X}^{n m} \, ,
\end{equation}
with $m = 1, \ldots , d-1$ and $n = 0, \ldots , d-1$. They fulfill
the pairwise orthogonality relations
\begin{eqnarray}
  \label{pairwise}
  & \Tr [ \op{\tilde{\Lambda}} ( m ) \, \op{\tilde{\Lambda}}^\dagger ( m^\prime)  ] =
  d \, \delta_{m m^\prime} \, .  & \nonumber \\
  &  & \\
  & \Tr [ \op{\Lambda} (m, n) \, \op{\Lambda}^\dagger (m^\prime, n^\prime ) ] = 
  d \, \delta_{m m^\prime} \, \delta_{n n^\prime} \,  ,  & \nonumber 
\end{eqnarray}
which indicate that, for every value of $n$, we generate a maximal set of 
$d-1$ commuting operators and that all these classes are disjoint.  In 
addition,  the common eigenstates of each class $n$ form different sets of MUBs.

If one recalls that the finite Fourier transform $F$ is~\cite{Vourdas:2004}
\begin{equation}
  \label{FT1}
  \op{F} = \frac{1}{\sqrt{d}} \sum_{\ell, \ell^\prime = 0}^{d-1}
  \omega( \ell \, \ell^\prime ) \, | \ell \rangle \langle \ell^\prime | \, ,
\end{equation}
then one easily verifies that
\begin{equation}
\op{Z} = \op{F} \, \op{X} \, \op{F}^\dagger \, ,
\end{equation}
much in the spirit of the standard way of looking at complementary
variables in the infinite-dimensional Hilbert space: the position and
momentum eigenstates are Fourier transform one of the other.  

The operators  $\op{\Lambda} (m, n)$ can be written as
\begin{equation}
\label{eq:Vact}
  \op{\Lambda} ( m, n ) =  e^{i \phi (m, n)} \,
  \op{V}^{n} \, \op{Z}^{m} \, \op{V}^{\dagger n} \, , 
\end{equation}
where $\op{V}$  turns out to be  ($d > 2$)
\begin{equation}
  \op{V} = \sum_{\ell = 0}^{d-1} \omega ( - 2^{-1} \ell^{2} ) \, 
  |\widetilde{\ell} \rangle \langle  \widetilde{\ell}| \, , 
\end{equation}
and the phase $\phi (m, n)$ is~\cite{Klimov:2006,Bjork:2008}
\begin{equation}
  \label{eq:7}
  \phi (m, n) = \omega ( 2^{-1} n m^{2}) \, .
\end{equation} 
Here $2^{-1}$ denotes the multiplicative inverse of 2 modulo $d$ [that
is, $2^{-1} = (d+1)/2$] and $| \widetilde{\ell} \rangle$ is the
conjugate basis, which is defined by the action of the Fourier
transform on the computational basis, namely $|\widetilde{\ell}
\rangle = \op{F} \, | \ell \rangle$.

The case of qubits ($d = 2$) requires minor modifications: 
$\op{V}$ is now
\begin{equation}
  \op{V} = \frac{1}{2} 
  \left ( 
    \begin{array}{cc}
      1+i \; & 1-i \\ 
      1-i \; & 1+i
    \end{array}
  \right )  \, ,
\end{equation}
while its action reads as $\op{V} \, \op{Z} \, \op{V}^{\dagger} = - i
\op{Z} \op{X}$.

The operator $\op{V}$ has quite an important property: its powers
generate MUBs when acting on the computational basis: indeed, if
\begin{equation}
  | \psi_{\ell}^{n} \rangle =  \op{V}^{n}  | \ell \rangle \, ,
\end{equation}
one can check by a direct calculation that the states $ 
| \psi_{\ell}^{n} \rangle$ fulfill (\ref{eq:MUB3}), which confirms the
unbiasedness. If we denote $\op{\Lambda}_{\ell \ell^\prime} (m,n) =
\langle \ell | \op{\Lambda} (m, n ) | \ell^\prime \rangle$, according
to Eq.~(\ref{eq:Vact}), we have
\begin{equation}
  \op{\Lambda}_{\ell \ell^\prime} (m,n) = e^{i\phi (m,n)} \, 
  \langle \psi _{c}^{n}|\op{Z}^{m}|\psi _{d}^{n}\rangle \, .
\end{equation}
Therefore, up to an unessential phase factor, 
$\op{\Lambda}_{\ell  \ell^\prime} (m,n)$ are the matrix elements of the 
powers of the diagonal operator $\op{Z}$ in the corresponding MUB. 
This provides an elegant interpretation of these objects, which will 
play an essential role in what follows.

\subsection{Qudit Bell states}

For the case of two qudits, a sensible generalization of Bell states
was devised in Ref.~\cite{Bennett:1993}, namely
\begin{equation}
  \label{eq:Bellmn}
  | \Psi_{m n} \rangle =  \frac{1}{\sqrt{d}} 
  \sum_{\ell = 0}^{d-1} \omega ( m \ell) \,
  | \ell \rangle_{A}  | \ell + n \rangle_{B}  \, ,
\end{equation}
where, to simplify as much as possible the notation, we drop the subscript 
$AB$ from $| \Psi_{mn} \rangle$, since we deal only with bipartite states.  
For further use, we also define
\begin{equation}
  \label{eq:Bellm}
  | \tilde{\Psi}_{m} \rangle =  \frac{1}{\sqrt{d}} 
  \sum_{\ell = 0}^{d-1} 
  | \ell \rangle_{A}  |\ell + m \rangle_{B}  \, .
\end{equation}
In the same vein, some generalized gates have been proposed to create
these $d^2$ states~\cite{Alber:2001,Durt:2003}.

This set of states is orthonormal 
\begin{eqnarray}
\label{eq:Bellort}
  & \langle \Psi_{m n}  | \Psi_{m^\prime n^\prime}  \rangle =
  \delta_{m m^\prime} \, \delta_{n n^\prime},
  \qquad 
  \langle \tilde{\Psi}_{m} | \tilde{\Psi}_{m^\prime} \rangle =
  \delta_{m m^\prime} \, , & \nonumber \\
  & & \\
  & \langle \Psi_{m n} | \tilde{\Psi}_{m^\prime} \rangle =
  \delta_{m 0} \, \delta_{m^\prime 0} \, , & 
  \nonumber  
\end{eqnarray}
and allows for a resolution of the identity
\begin{equation}
  \label{eq:Bellid}
  \sum_{m=1}^{d-1} \sum_{n=0}^{d-1} 
  |\Psi_{m n} \rangle  \langle \Psi_{m n} | + 
  \sum_{m=1}^{d-1} 
  |\tilde{\Psi}_{m} \rangle \langle  \tilde{\Psi}_{m}| 
  = \openone \, , 
\end{equation}
so they constitute a \textit{bona fide} basis for any bipartite qudit
system.  As anticipated in the Introduction, there must be then a
connection with MUBs.  And this is indeed the case: it suffices to
observe that the states (\ref{eq:Bellmn}) and (\ref{eq:Bellm}) can be
recast as
\begin{eqnarray}
  \label{eq:BellLam}
  & \displaystyle
  | \Psi_{m n} \rangle = \frac{1}{\sqrt{d}} 
  \sum_{\ell, \ell^\prime = 0}^{d-1} \op{\Lambda}_{\ell \ell^\prime} (m,n) \,
  | \ell \rangle_{A} |\ell^\prime \rangle_{B} \, , & \nonumber \\
  &  & \\
  &  \displaystyle
  | \tilde{\Psi}_{m} \rangle = \frac{1}{\sqrt{d}} 
  \sum_{\ell, \ell^\prime = 0}^{d-1} \tilde{\Lambda}_{\ell \ell^\prime} (m) \,
  |\ell \rangle_{A} |\ell^\prime \rangle_{B} \, , & \nonumber
\end{eqnarray}
which can be checked by a direct calculation and
 $\op{\Lambda}_{\ell  \ell^\prime} (m,n)$ and 
 $ \tilde{\Lambda}_{\ell \ell^\prime} (m)$
are the matrix elements of the operators (\ref{SCop}).

The matrices $\Lambda $ possess quite an interesting symmetry 
property
\begin{equation}
  \label{SR1}
  \op{\Lambda}_{\ell \ell^\prime} (m, n) = \omega( m^{2} n ) \, 
  \op{\Lambda}_{\ell^\prime \ell} (m, n)  \, , 
  \quad 
  \tilde{\Lambda}_{\ell \ell^\prime} ( m ) =
  \tilde{\Lambda}_{\ell^\prime \ell} (m) \, . 
\end{equation}
In consequence, $\tilde{\Lambda} (m) $ are always totally symmetric
under the permutation of subsystems $A$ and $B$ and so are the
corresponding Bell states. Whenever $\omega( m^{2} n ) = \pm 1$,
$\op{\Lambda} (m, n)$ are either symmetric or antisymmetric.  This
happens for $m n = 0$ $\pmod{d}$, and this is only possible for
qubits: the symmetric matrices are $\tilde{\Lambda} ( 0 )$,
$\tilde{\Lambda} (1)$, and $\op{\Lambda} (1, 0)$, while the
antisymmetric is $\Lambda (1, 1)$.  The corresponding symmetric states
are $| \tilde{\Psi}_0 \rangle = | \Phi_{+}\rangle$, $| \tilde{\Psi}_1
\rangle = | \Psi_{+} \rangle$, and $ | \Psi_{1,0} \rangle = | \Phi_{-}
\rangle$, and $ |\Psi_{1,1} \rangle=|\Psi _{-}\rangle$ is the
antisymmetric one.

Finally, we can sum up the projectors of the bipartite states
(\ref{eq:Bellmn}) over $m$, obtaining the following interesting
novel property:
\begin{eqnarray}
  \label{Theo}
  & \displaystyle
  \sum_{m=0}^{d-1} | \Psi_{mn} \rangle \langle \Psi_{mn} |  = 
  \frac{1}{d} \sum_{\ell = 0}^{d-1} (\op{X}^{n \ell} \op{Z}^{-\ell} )_{A} 
  \otimes (\op{X}^{n \ell}Z^{\ell} )_{B} \, , & \nonumber \\
  & & \\
  & \displaystyle
  \sum_{m=0}^{d-1} | \tilde{\Psi}_{m} \rangle \langle \tilde{\Psi}_{m}|  
  =  \frac{1}{d} \sum_{\ell = 0}^{d-1} ( \op{X}^{\ell})_{A}\otimes 
  (\op{X}^{\ell})_{B} \, . & \nonumber
\end{eqnarray}
In words, this means that the sum of projectors over the index $m$ 
is the sum of direct product of commuting operators for each
particle. The proof of this statement involves a tedious yet direct
calculation.

For the case of two qubits, this implies that
\begin{eqnarray}
  & \displaystyle
  \sum_{m=0,1} | \Psi_{m1} \rangle \langle \Psi_{m1} | = \frac{1}{2}
  [ \openone + (\op{X} \op{Z})_{A} \otimes (\op{X} \op{Z})_{B} ] \, , & 
  \nonumber \\
  & & \\
  &  \displaystyle
  \sum_{m=0,1} | \tilde{\Psi}_{m} \rangle \langle \tilde{\Psi}_{m} | = 
  \frac{1}{2} [ \openone + (\op{X} )_{A} \otimes (\op{X} )_{B} ] \, . &
  \nonumber
\end{eqnarray}

\section{Bipartite multiqudit systems}

\subsection{Mutually unbiased bases for $n$ qudits}

The previous ideas can be extended for a system of $n$ qudits.
Instead of natural numbers, it is then convenient to use elements of
the finite field $\Gal{d^{n}}$ to label states, since then we can
almost directly translate all the properties studied before for a
single qudit.  In the Appendix we briefly summarize the basic notions
of finite fields needed to proceed.

We denote as $ | \lambda \rangle$ (from here on, Greek letters
will represent elements in the field $ \Gal{d^{n}}$) an orthonormal
basis in the Hilbert space of the quantum system. Operationally, the
elements of the basis can be labelled by powers of the primitive element,
which can be found as roots of a minimal irreducible polynomial of
degree $n$ over $\mathbb{Z}_{d}$.

The generators of the generalized Pauli group are now
\begin{equation}
  \op{X}_{\mu} | \lambda \rangle = | \lambda + \mu \rangle \, ,
  \qquad
  \op{Z}_{\mu} | \lambda  \rangle = \chi ( \lambda \mu ) | \lambda \rangle \, ,
\end{equation}
where $\chi (\lambda )$ is an additive character (defined in the
Appendix). The Weyl form of the commutation relations reads as $
\op{Z}_{\mu} \op{X}_{\nu} = \chi ( \mu \nu ) \op{X}_{\nu}  \op{Z}_{\mu}$.

In agreement with (\ref{SCop}), we introduce the set of monomials
\begin{equation}
  \label{set1}
  \tilde{\Lambda} ( \mu ) =  \op{X}_{\mu} \, , 
  \qquad
  \op{\Lambda} (\mu, \nu) = \op{Z}_{\mu} \op{X}_{\nu \mu} \, , 
\end{equation}
and their corresponding eigenstates also form a complete set of 
$d^{n}+1$ MUBs.

The finite Fourier transform now is~\cite{Vourdas:2005}
\begin{equation}
  \op{F} = \frac{1}{\sqrt{d^n}}  \sum_{\lambda ,\lambda^\prime \in} 
  \chi ( \lambda \, \lambda^\prime ) |  \lambda \rangle 
  \langle \lambda^\prime | \, ,
  \label{03_}
\end{equation}
and thus
\begin{equation}
  \op{Z}_{\mu} = \op{F} \, \op{X}_{\mu} \, \op{F}^{\dagger} \, .
  \label{ZFX}
\end{equation}

The rotation operator $\op{V}_{\nu}$ transforms the diagonal
$\op{Z}_{\mu}$ into an arbitrary monomial according to
\begin{equation}
  \label{s1}
  \op{\Lambda} (\mu, \nu )   = e^{i \varphi ( \mu , \nu )} \, 
  \op{V}_{\nu} \, \op{Z}_{\alpha} \, \op{V}_{\nu}^\dagger \, , 
\end{equation}
and is diagonal in the conjugate basis (defined, as before,
via the Fourier transform $|\widetilde{\lambda} \rangle =
\op{F} \, | \lambda \rangle$)
\begin{equation}
  \label{V}
  V_{\nu} = \sum_{\lambda} c_{\lambda \nu} \,
  |\widetilde{\lambda} \rangle \langle \widetilde{\lambda}| \, , 
\end{equation}
where the coefficients $c_{\lambda \nu}$ satisfy the following relation
\begin{equation}
  \label{ck_2}
  c_{0 \nu} = 1 \, , \qquad
  c_{\lambda + \lambda^\prime \, \nu} \, c_{\lambda \nu}^{\ast} = 
  c_{\lambda^\prime \nu} \chi ( - \nu \lambda^\prime \lambda ) , 
\end{equation}
When $ d \neq 2$, a particular solution of Eq.~(\ref{ck_2}) is
\begin{equation}
  \label{ck_p}
  c_{\lambda  \nu} = \chi ( - 2^{-1 } \lambda^{2} \nu ) .
\end{equation}

Again, if we define the states
\begin{equation}
  | \psi_{\lambda}^{\mu} \rangle = \op{V}_{\mu} |\lambda \rangle \, ,  
  \label{psi_1}
\end{equation}
they are unbiased and $\Lambda_{\lambda \lambda^\prime} (\mu ,\nu)$
are the matrix elements of the diagonal operator $Z_{\mu}$ on the
corresponding MUB
\begin{equation}
  \Lambda_{\lambda \lambda^\prime} (\mu ,\nu ) = e^{i\varphi ( \mu ,\nu )} 
  \langle \psi_{\lambda}^{\nu}|Z_{\mu}|\psi_{\lambda^\prime}^{\mu} \rangle \, . 
\end{equation}

\subsection{Multiqudit Bell states}

For a bipartite system of $n$ qudits, it seems natural to extend the
previous construction (\ref{eq:BellLam}) by introducing the $d^{2n}$
states
\begin{eqnarray}
  |\Psi_{\mu \nu} \rangle & = & \frac{1}{\sqrt{d^n}} 
  \sum_{\lambda, \lambda^\prime} \Lambda _{\lambda \lambda^\prime} (\mu ,\nu ) \,
  |\lambda \rangle_{A} |\lambda^\prime \rangle_{B} \, , \nonumber  \\
  & & \\
  |\tilde{\Psi}_\mu \rangle & = & \frac{1}{\sqrt{d^n}}
  \sum_{\lambda ,\lambda^\prime} \tilde{\Lambda}_{\lambda \lambda^\prime} (\mu ) \,
  |\lambda \rangle_{A} |\lambda^\prime \rangle_{B} \, . \nonumber
\end{eqnarray}

Accordingly, the associated Bell states are (apart from an unessential
global phase)
\begin{eqnarray}
  |\Psi_{\mu \nu} \rangle & = &\frac{1}{\sqrt{d^n}} \sum_{\lambda} 
  \chi (\mu \lambda) \, | \lambda \rangle_{A} |\lambda + \nu \rangle_{B} \, ,
  \nonumber \\
  & & \\
  |\tilde{\Psi}_{\mu} \rangle & = & \frac{1}{\sqrt{d^n}} \sum_{\lambda} 
  |\lambda  \rangle_{A} |\lambda + \nu \rangle_{B} \, , \nonumber 
\end{eqnarray}
which look as quite a reasonable generalization.  One can prove the
orthogonality
\begin{eqnarray}
  \label{eq:Bellortf}
  & \langle \Psi_{\mu \nu}  | \Psi_{\mu^\prime \nu^\prime}  \rangle =
  \delta_{\mu \mu^\prime} \, \delta_{\nu \nu^\prime},
  \qquad 
  \langle \tilde{\Psi}_{\mu} | \tilde{\Psi}_{\mu^\prime} \rangle =
  \delta_{\mu \mu^\prime} \, , & \nonumber \\
  & & \\
  & \langle \Psi_{\mu \nu} | \tilde{\Psi}_{\mu^\prime} \rangle =
  \delta_{\mu 0} \, \delta_{\mu^\prime 0} \, ,  & 
  \nonumber  
\end{eqnarray}
and the completeness relation
\begin{equation}
  \sum_{\mu \neq 0, \nu} | \Psi_{\mu \nu} \rangle \langle \Psi_{\mu \nu} | +
  \sum_{\mu} | \tilde{\Psi}_{\mu} \rangle \langle \tilde{\Psi}_{\mu}| = 
  \openone \, , 
\end{equation}
which confirms that they constitute a basis.  Moreover, the reduced
density matrices for both subsystems are completely random
\begin{equation}
  \Tr_{A} ( | \Psi_{\mu \nu} \rangle  \langle \Psi_{\mu \nu} )  = 
  \frac{1}{d^n} \sum_{\lambda} | \lambda \rangle_{B} \, \, 
  {}_{B} \langle \lambda | \, , 
\end{equation}
(and other equivalent equation with $A$ and $B$ interchanged) showing
that they are maximally entangled states.

The concept of symmetric and antisymmetric states can be worked
out for systems of $n$ qubits, which constitutes a nontrivial 
generalization of our previous discussion~\cite{Sych:2009,Jex:2003}.
 The symmetric states [i.e.,
$\Lambda_{\lambda \lambda^\prime} (\mu, \nu) = \Lambda_{\lambda^\prime
  \lambda} (\mu, \nu)]$, correspond to those pairs $(\mu, \nu )$ such
that
\begin{equation}
  \tr ( \nu \mu^{2} ) = 0 \, ,
\end{equation}
where $\tr$, in small case, denotes the trace map in the field.
Clearly, all the states $|\Psi_{\mu 0}\rangle $ and
$|\tilde{\Psi}_{\mu}\rangle $ are symmetric. The antisymmetric states
[i.e., $\Lambda_{\lambda \lambda^\prime} (\mu, \nu) = -
\Lambda_{\lambda^\prime \lambda} (\mu, \nu)]$ are defined by the pairs
$(\mu ,\nu )$ such that
\begin{equation}
  \tr ( \nu \mu^{2} ) = 1 \, .
\end{equation}

Finally, a property similar to (\ref{Theo}) is fulfilled: summing up
the projectors over $\mu$ one obtains
\begin{eqnarray}
  & \displaystyle
  \sum_{\mu}| \Psi_{\mu \nu} \rangle \langle \Psi_{\mu \nu} | = 
  \sum_{\lambda} (X_{\lambda \nu} Z_{-\lambda})_{A} \otimes 
  (X_{\lambda \nu}Z_{\lambda})_{B} \, , &
  \nonumber \\
  & & \\
  & \displaystyle
  \sum_{\mu}|\tilde{\Psi}_{\mu} \rangle \langle \tilde{\Psi}_{\mu} | =
  \sum_{\lambda} (X_{\lambda})_{A} \otimes (X_{\lambda})_{B} \, , &  
  \nonumber
\end{eqnarray}
whose interpretation is otherwise the same as for qudits.

\subsection{Examples}

Since we are dealing with $n$-qudit systems, we can map the abstract
Hilbert space $\mathcal{H}_{d^n}$ into $n$ single-qudit Hilbert spaces.
This is achieved by expanding any field element in a convenient
basis $\{ \theta_{j} \}$ (with $j = 1, \ldots, n$), so that
\begin{equation}
   \lambda  = \sum_{j} \ell_{j} \, \theta_{j} \, , 
\end{equation}
where $\ell_{j} \in \mathbb{Z}_{d}$. Then, we can represent the states
as $ | \lambda \rangle = | \ell_{1}, \ldots, \ell_{n} \rangle$ and the
coefficients $\ell_{j}$ play the role of quantum numbers for each
qudit. 

For example, for two qubits, the abstract state $( |0 \rangle + 
| \sigma^3 \rangle)/\sqrt{2}$, where $\sigma$ is a primitive
elements, can be mapped onto the physical state $ | 0 0 \rangle + 
| 1 0 \rangle )/ \sqrt{2}$ in the polynomial basis $\{ 1, \sigma \}$,
whereas in the selfdual basis $\{ \sigma, \sigma^2 \}$ it is
associated with $( |00 \rangle + | 11 \rangle )/\sqrt{2}$. Observe
that, while the first state is factorizable, the other one is
entangled.

The use of the selfdual basis (or the almost selfdual, if
the latter does not exist) is especially advantageous, since only
then the  Fourier transform and the basic operators factorize in 
terms of single-qudit analogues:
\begin{equation}
  \op{X}_{\lambda} = \op{X}^{\ell_{1}} \otimes \ldots 
  \otimes \op{X}^{\ell_{n}} \, ,
  \qquad
  \op{Z}_{\lambda} = \op{Z}^{\ell_{1}} \otimes \ldots 
  \otimes \op{Z}^{\ell_{n}} \, . 
\end{equation}

For a bipartite $4 \times 4$ system the states are represented as 
$| \lambda \rangle = |\ell_{1}, \ell_{2} \rangle$ with $\ell_{j} \in \mathbb{Z}_{2}$.  The Bell basis can be expressed as
\begin{widetext}
  \begin{eqnarray}
    | m_{1}, n_{1}; m_{2},n_{2} \rangle & = &
    \frac{(-1)^{m_{1} n_{2} + m_{2} n_{1}}}{2}
    \sum_{\ell_{1},\ell_{2}} (-1)^{m_{1} \ell_{1} + m_{2} \ell_{2}} \, 
    |\ell_{1} + m_{1} n_{2} + m_{2} n_{1}, 
    \ell_{2} + m_{1} n_{1} + m_{2} n_{2} \rangle_{A}
    |\ell_{1}, \ell_{2} \rangle_{B} \, , \nonumber \\
    & & \\
    |\widetilde{m_{1}, m_{2}} \rangle & = & \frac{1}{2}
    \sum_{\ell_{1}, \ell_{2}} |\ell_{1} + n_{1}, \ell_{2} + m_{2} \rangle_{A} 
    |\ell_{1}; \ell_{2}\rangle_{B} \, . \nonumber 
  \end{eqnarray}
\end{widetext}
The conditions
\begin{equation}
  \label{eq:1}
  m_{1} n_{2} + m_{2} n_{1} = \left \{
\begin{array}{l} 
0 \, ,\\ 
1 \, ,
\end{array}
\right .
\end{equation}
determine the symmetric and antisymmetric states, respectively.  The
solutions of this equation show that there are 10 symmetric states and
6 antisymmetric ones, whose explicit form can be computed from
previous formulas.

Before ending, we wish to stress that so far we have been dealing with
systems made of $n$ qudits. However, sometimes they can be treated
instead as a single `particle' with $d^n$ levels.  For example, a
four-dimensional system can be taken as two qubits or as a ququart.
If, for some physical reason, we choose for the quqart, we can still
use Eq.~(\ref{eq:Bellmn}), as in Ref.~\cite{Bennett:1993}, even if now
the dimension is not a prime number. However, if we proceed in this
way the resulting basis contains 6 symmetric and 2 antisymmetric
states, while the other 8 do not have any explicit symmetry, 
contrary to our results.

\section{Concluding remarks}

In summary, we have provided a complete MUB-based construction of Bell 
states that fulfills all the requirements needed for a good description of maximally entangled states of bipartite multiqudit systems.

Mutually unbiasedness is a very deep concept arising from the exact
formulation of complementarity. The deep connection shown in this
paper with Bell bases is more than a mere academic curiosity, for it
is immediately applicable to a variety of experiments involving qudit
systems.

\appendix

\section{Finite fields}
\label{Sec: Galois}

In this appendix we briefly recall the minimum background needed in
this paper.  The reader interested in more mathematical details is
referred, e.g., to the excellent monograph by Lidl and
Niederreiter~\cite{Lidl:1986}.

A commutative ring is a nonempty set $R$ furnished with two binary
operations, called addition and multiplication, such that it is an
Abelian group with respect the addition, and the multiplication is
associative.  Perhaps, the motivating example is the ring of integers
$\mathbb{Z}$, with the standard sum and multiplication. On the other
hand, the simplest example of a finite ring is the set $\mathbb{Z}_n$
of integers modulo $n$, which has exactly $n$ elements.

A field $F$ is a commutative ring with division, that is, such that 0
does not equal 1 and all elements of $F$ except 0 have a multiplicative 
inverse (note that 0 and 1 here stand for the identity elements for the 
addition and multiplication, respectively, which may differ from the 
familiar real numbers 0 and 1). Elements of a field form Abelian 
groups with respect to addition and multiplication (in this latter
case, the zero element is excluded).

The characteristic of a finite field is the smallest integer $d$ such 
that
\begin{equation}
d \, 1= \underbrace{1 + 1 + \ldots + 1}_{\mbox{\scriptsize $d$ times}}=0
\end{equation}
and it is always a prime number. Any finite field contains a prime 
subfield $\mathbb{Z}_d$ and has $d^n$ elements, where $n$ is a 
natural number. Moreover, the finite field containing $d^{n}$ elements 
is unique and is called the Galois field $\Gal{d^n}$.

Let us denote as $\mathbb{Z}_{d} [x]$ the ring of polynomials with
coefficients in $\mathbb{Z}_{d}$. Let $P(x)$ be an irreducible
polynomial of degree $n$ (i.e., one that cannot be factorized over
$\mathbb{Z}_{d}$). Then, the quotient space $\mathbb{Z}_{d}[X]/P(x)$
provides an adequate representation of $\Gal{d^n}$. Its elements can 
be written as polynomials that are defined modulo the irreducible 
polynomial $P(x)$. The multiplicative group of $\Gal{d^n}$ is cyclic
and its generator is called a primitive element of the field.

As a simple example of a nonprime field, we consider the 
polynomial $x^{2}+x+1=0$, which is irreducible in $\mathbb{Z}_{2}$.
If $\prim$ is a root of this polynomial, the elements $\{ 0, 1, \prim , 
\prim^{2} = \prim + 1 = \prim^{-1} \} $ form the finite field $\Gal{2^2}$ 
and $\prim$ is a primitive element.

A basic map is the trace
\begin{equation}
 \label{deftr}
 \tr (\lambda ) = \lambda + \lambda^{2} + \ldots +
 \lambda^{d^{n-1}} \, .
\end{equation}
It is always in the prime field $\mathbb{Z}_d$ and satisfies
\begin{equation}
 \label{tracesum}
 \tr ( \lambda + \lambda^\prime ) =
 \tr ( \lambda ) + \tr ( \lambda^\prime ) \, .
\end{equation}
In terms of it we define the additive characters as
\begin{equation}
 \label{Eq: addchardef}
 \chi (\lambda ) = \exp \left [ \frac{2 \pi i}{p}
  \tr ( \lambda ) \right] \, ,
\end{equation}
which posses two important properties:
\begin{equation}
 \chi (\lambda + \lambda^\prime ) =
 \chi (\lambda ) \chi ( \lambda^\prime ) ,
 \qquad
 \sum_{\lambda^\prime \in
  \Gal{d^n}} \chi ( \lambda \lambda^\prime ) = d^n
 \delta_{0,\lambda} \, .
 \label{eq:addcharprop}
\end{equation}

Any finite field $\Gal{d^n}$ can be also considered as an
$n$-dimensional linear vector space. Given a basis $\{ \theta_{j} \}$,
($j = 1,\ldots, n$) in this vector space, any field element can be
represented as
\begin{equation}
  \label{mapnum}
  \lambda = \sum_{j=1}^{n} \ell_{j} \, \theta_{j} ,
\end{equation}
with $\ell_{j}\in \mathbb{Z}_{d}$. In this way, we map each element of
$\Gal{d^n}$ onto an ordered set of natural numbers $\lambda
\Leftrightarrow (\ell_{1}, \ldots , \ell_{n})$.

Two bases $\{ \theta_{1}, \ldots, \theta_{n} \} $ and $\{
\theta_{1}^\prime, \ldots , \theta_{n}^\prime \} $ are dual when
\begin{equation}
  \tr ( \theta_{k} \theta_{l}^\prime ) =\delta_{k,l}.
\end{equation}
A basis that is dual to itself is called selfdual.

There are several natural bases in $\Gal{d^n}$. One is the polynomial
basis, defined as
\begin{equation}
 \label{polynomial}
 \{1, \prim, \prim^{2}, \ldots, \prim^{n-1} \} ,
\end{equation}
where $\prim $ is a primitive element. An alternative is the normal
basis, constituted of
\begin{equation}
 \label{normal}
 \{\prim, \prim^{d}, \ldots, \prim^{d^{n-1}} \}.
\end{equation}
The choice of the appropriate basis depends on the specific problem 
at hand. For example, in $\Gal{2^2}$ the elements $\{ \prim , \prim^{2}\}$ 
are both roots of the irreducible polynomial. The polynomial basis 
is $\{ 1, \prim \} $ and its dual is $\{ \prim^{2}, 1 \}$, while 
the normal basis $\{ \prim , \prim^{2} \} $ is selfdual.
  
The selfdual basis exists if and only if either $d$ is even or both
$n$ and $d$ are odd. However for every prime power $d^n$, there exists
an almost selfdual basis of $\Gal{d^n}$, which satisfies the
properties: $\tr ( \theta_{i} \theta_{j} ) =0$ when $i\neq j$ and $\tr
( \theta_{i}^{2} ) =1$, with one possible exception. For instance, in
the case of two qutrits $\Gal{3^2}$, a selfdual basis does not exist
and two elements $\{ \prim^2, \prim^4\}$, $\prim$ being a root of the
irreducible polynomial $x^2 + x + 2 =0$, form a self dual basis
\begin{equation}
  \label{eq:2}
  \tr (\prim^2 \prim^2) = 1 \, , \quad
  \tr (\prim^4 \prim^4) = 2 \, , \quad 
  \tr (\prim^2 \prim^4) = 0 \, . 
\end{equation}


\begin{thebibliography}{99}
\bibitem{Schrodinger:1935}
E. Schr\"{o}dinger, 
Math. Proc. Cambridge Philos. Soc. \textbf{31}, 555 (1935).

\bibitem{Nielsen:2000}
M. A. Nielsen and I. L. Chuang, 
\textit{Quantum Computation and Quantum Information} 
(Cambridge University Press, Cambridge, 2000).

\bibitem{Peres:1993}
A. Peres,
\textit{Quantum Theory: Concepts and Methods}
(Kluwer Academic, Boston, 1993).

\bibitem{Wei:2007}
T. C. Wei, J. T. Barreiro, and P. G. Kwiat,
Phys. Rev. A \textbf{75}, 060305(R) (2007).

\bibitem{Dur:2000}
W. D\"{u}r and J. I. Cirac, 
Phys. Rev. A \textbf{61}, 042314 (2000). 

\bibitem{Durr:2000b}
W. D\"{u}r, G. Vidal, and J. I. Cirac, 
Phys. Rev. A \textbf{62}, 062314 (2000).

\bibitem{Acin:2000}
A.~Ac\'{\i}n, A.~Andrianov, L.~Costa, E.~Jan\'{e}, 
J.~I.~Latorre, and R.~Tarrach,
Phys. Rev. Lett. \textbf{85}, 1560 (2000).

\bibitem{Briegel:2001}
H. J. Briegel and R. Raussendorf,
Phys. Rev. Lett. \textbf{86}, 910 (2001).

\bibitem{Verstraete:2002}
F. Verstraete, J. Dehaene, B. DeMoor, and H. Verschelde, 
Phys. Rev. A \textbf{65}, 052112 (2002).

\bibitem{Rigolin:2006}
G. Rigolin, T. R. de Oliveira, and M. C. de Oliveira,
Phys. Rev. A \textbf{74}, 022314 (2006).

\bibitem{Facchi:2008}
P. Facchi, G. Florio, G. Parisi, and S. Pascazio,
Phys. Rev. A \textbf{77}, 060304 (2008).

\bibitem{Bechmann:2000} 
H. Bechmann-Pasquinucci and W. Tittel, 
Phys. Rev. A \textbf{61}, 062308 (2000).

\bibitem{Bourennane:2001}
M. Bourennane, A. Karlsson,  and G. Bj\"{o}rk,
 Phys. Rev. A \textbf{64}, 012306 (2001).
 
\bibitem{Cerf:2002}
N. J. Cerf, M. Bourennane, A. Karlsson, and N. Gisin,
Phys. Rev. Lett. \textbf{88}, 127902 (2002).

\bibitem{Sych:2004}
D. Sych, B. Grishanin, and V. Zadkov, 
Phys. Rev. A \textbf{70}, 052331 (2004).

\bibitem{Sych:2009} 
D. Sych and G. Leuchs,
New J. Phys. \textbf{11}, 013006 (2009).

\bibitem{Schwinger:1960}
J. Schwinger, 
Proc. Natl. Acad. Sci. USA \textbf{46}, 570 (1960).

\bibitem{Wootters:1987}
W. K. Wootters,
Ann. Phys. (NY) \textbf{176}, 1 (1987).

\bibitem{Wootters:1989}
W. K. Wootters and B. D. Fields,
Ann. Phys. (NY) \textbf{191}, 363 (1989).

\bibitem{Wootters:2004}
W. K. Wootters,
IBM J. Res. Dev. \textbf{48}, 99 (2004).

\bibitem{Gibbons:2004b}
K. S. Gibbons, M. J. Hoffman, and W. K. Wootters,
Phys. Rev. A \textbf{70}, 062101 (2004).

\bibitem{Wootters:2006}
W. K. Wootters, 
Found. Phys. \textbf{36}, 112 (2006).

\bibitem{Planat:2005}
M. Planat and H. C. Rosu, 
Eur. Phys. J. D \textbf{36}, 133 (2005).

\bibitem{Bengtsson:2007}
I. Bengtsson, W. Bruzda, \AA. Ericsson, J. \AA. Larsson,
W. Tadej, and K. \.Zyczkowski,
J. Math. Phys. \textbf{48}, 052106 (2007).

\bibitem{Asplund:2001}
R. Asplund, G. Bj\"ork, and M. Bourennane,
J. Opt. B \textbf{3}, 163 (2001).

\bibitem{Gottesman:1996}
D. Gottesman, 
Phys. Rev. A \textbf{54}, 1862 (1996).

\bibitem{Calderbank:1997}
A. R. Calderbank, E. M. Rains, P. W. Shor, and N. J. A. Sloane,
Phys. Rev. Lett. \textbf{78}, 405 (1997).

\bibitem{Englert:2001}
B.-G. Englert and Y. Aharonov,
Phys. Lett. A \textbf{284}, 1 (2001).

\bibitem{Aravind:2003}
P. K. Aravind, 
Z. Naturforsch. A: Phys. Sci. \textbf{58}, 2212 (2003).

\bibitem{Paz:2005}
J. P. Paz, A. J. Roncaglia, and M. Saraceno, 
Phys. Rev. A \textbf{72}, 012309 (2005).

\bibitem{Kimura:2006}
G. Kimura, H. Tanaka, and M.  Ozawa
Phys. Rev. A \textbf{73} 050301(R) (2006).

\bibitem{Ivanovic:1981}
I. D. Ivanovi\'{c},
J. Phys. A \textbf{14}, 3241 (1981).

\bibitem{Kraus:1987}
K.~Kraus, 
Phys. Rev. D \textbf{35}, 3070 (1987).

\bibitem{Lawrence:2002}
J.~Lawrence, \v{C}. Brukner, A.~Zeilinger, 
Phys. Rev. A \textbf{65} 032320 (2002).

\bibitem{Bandyopadhyay:2002}
S.~Bandyopadhyay, P.~O.~Boykin, V.~Roychowdhury, and V.~Vatan, 
Algorithmica \textbf{34}, 512 (2002).

\bibitem{Klappenecker:2004}
A. Klappenecker and M. R\"otteler, 
Lecture Notes in Comput. Sci. \textbf{2948}, 137 (2004).

\bibitem{Lawrence:2004}
J. Lawrence,
Phys. Rev. A \textbf{70}, 012302 (2004).

\bibitem{Pittenger:2005}
A. O. Pittenger and M. H. Rubin,
J. Phys. A \textbf{38}, 6005 (2005).

\bibitem{Wocjan:2005}
P. Wocjan and T. Beth,
Quantum Inform. Compu. \textbf{5}, 93 (2005).

\bibitem{Durt:2005}
T. Durt,  
J. Phys. A \textbf{38}, 5267 (2005).

\bibitem{Klimov:2007}
A. B. Klimov, J. L. Romero, G. Bj\"{o}rk, and 
L. L. S\'anchez-Soto,
J. Phys. A \textbf{40}, 3987 (2007).

\bibitem{Klimov:2005}
A. B. Klimov, L. L. S\'{a}nchez-Soto, and H. de Guise, 2005,
J. Phys. A \textbf{38}, 2747 (2005).

\bibitem{Weil:1964}
A. Weil,
Acta Math. \textbf{111}, 143 (1964).

\bibitem{Putnam:1987}
C. P. Putnam,
\textit{Commutation Properties of Hilbert Space Operators}
(Springer, Heidelberg, 1987).

\bibitem{Vourdas:2004}
A. Vourdas,
Rep. Prog. Phys. \textbf{67}, 267 (2004).

\bibitem{Klimov:2006}
A. B. Klimov, C. Mu\~{n}oz, and J. L. Romero,
J. Phys. A \textbf{39}, 14471 (2006).

\bibitem{Bjork:2008}
G. Bj\"{o}rk,, A. B. Klimov,  and 
L. L. S\'anchez-Soto,
Prog. Opt. \textbf{51}, 469 (2008).

\bibitem{Bennett:1993}
Ch. Bennett, G. Brassard, C. Cr\'epeau, R. Jozsa,
A. Peres, and W. K. Wootters,
Phys. Rev. Lett. \textbf{70}, 1895 (1993).

\bibitem{Alber:2001}
G. Alber, A. Delgado, N. Gisin, and I. Jex,
J. Phys. A \textbf{34}, 8821 (2001).

\bibitem{Durt:2003}
T. Durt and B. Nagler, 
Phys. Rev. A \textbf{68}, 042323 (2003). 

\bibitem{Vourdas:2005}
A. Vourdas, 
J. Phys. A \textbf{40}, R285 (2005).

\bibitem{Jex:2003}
I. Jex, G. Alber, S. M. Barnett, and A. Delgado,
Fortschr. Phys. \textbf{51}, 172 (2003). 

\bibitem{Lidl:1986}
R. Lidl and H. Niederreiter 
\textit{Introduction to Finite Fields and their Applications}
(Cambridge University Press, Cambridge, 1986).
\end{thebibliography}
\end{document}